\documentclass[showpacs,aps,graphicx,]{revtex4-2}
\usepackage{amsmath}
\usepackage{mathrsfs}
\usepackage{amsfonts}
\usepackage{amssymb}
\usepackage{graphicx}
\usepackage{caption}
\usepackage{subcaption} 
\usepackage{eufrak}
\usepackage{multirow}
\usepackage{float}
\usepackage[colorlinks,linkcolor=blue,anchorcolor=blue,citecolor=blue]{hyperref}
\usepackage{soul,color,xcolor}
\soulregister\upcite7
\soulregister\citep7
\soulregister\citet7
\soulregister\ref7
\soulregister\pageref7
\soulregister\and7
\soulregister\Name7

\begin{document}

\title{Entanglement purification for arbitrary multipartite high-dimensional Greenberger-Horne-Zeilinger state}

\author{Yu-Qing Kong, Jia-Ye Liu, Hui-Hui Kong, and Hai-Rui Wei\footnote{Corresponding author: hrwei@ustb.edu.cn} }

\address{ School of Mathematics and Physics, University of Science and Technology Beijing, Beijing 100083, China}

\date{\today }


%
%
%
%
%
%
%
%

\begin{abstract}
High-dimensional qudit (i.e., $d$-level or $d$-state) systems outperform two-dimensional qubit (i.e., 2-level or 2-state) systems in some quantum information processing tasks.
We exploit entanglement purification protocols (EPPs) for extracting a subset of high quality arbitrary $d$-dimensional $n$-partite Greenberger-Horne-Zeilinger (GHZ) states from a large set of less entangled GHZ states. 
In our protocols, qudit-flip and phase-flip errors can be corrected, and the fidelity of the output state can be asymptotically improved to unity by iterating the EPP process.  
Moreover, the schemes are immune to the number of polluted photons, 
the fidelity thresholds of the proposed EPPs are developed,
and the spatial-based single-qudit operations can be well manipulated with a range of balanced beam splitters, and phase shifters.  
These features make the proposed schemes offer an alternative method for high-dimensional multipartite entanglement purification.
\end{abstract}


\maketitle

\newcommand{\upcite}[1]{\textsuperscript{\textsuperscript{\cite{#1}}}}

\section{Introduction}\label{sec1}


The generation and distribution of high-fidelity maximally entangled states over long-distance nodes are essential for distributed quantum computation and large-scale quantum networks, such as quantum teleportation \cite{Teleportation1,Teleportation2,Teleportation3}, quantum cryptography \cite{cryptography1,cryptography2}, and quantum secure direct communication \cite{QSDC1,QSDC2,QSDC3,QSDC4}. However, in practical applications, the required pre-shared entangled states are inherently fragile. Their quality inevitably degrades due to environmental noise and channel losses, such as decoherence, dispersion, and attenuation, during entanglement distribution, which significantly hinders the performance of quantum information processing (QIP). To mitigate this environment-induced decoherence, several strategies have been developed, including entanglement purification protocols (EPPs) \cite{EPP1,EPP2,EPP3,EPP4,EPP5}, entanglement concentration protocols (ECPs) \cite{ECP1,ECP3,ECP4,ECP5}, quantum error correction codes (QECCs) \cite{QECCS}, quantum error-rejection codes (QERCs) \cite{QERJ}, decoherence-free spaces (DFSs) \cite{DFS1,DFS2}, and channel purifications (CPs) \cite{CP1,CP2,CP3}. Complex encoding and decoding operations are necessary for QECCs and QERCs, collective operations are required for CPs, and encoded entangled states are necessary for DFSs.

EPPs have emerged as a crucial technique for restoring entanglement quality from mixed states using local operations and classical communication (LOCC). By contrast, ECPs aim to extract maximally entangled states from a set of less-entangled pure states \cite{ECP6,ECP7,ECP8,ECP9}. Moreover, the error tolerance threshold of EPPs is generally higher than that of QECCs, whereas the requirements of QECCs are usually more stringent. The existing EPPs can be divided into several groups. In the first group, EPPs are accomplished with the help of controlled-NOT (CNOT) gates or similar bipartite entangled operations \cite{EPP-CNOT}. In the second group, the role of CNOT gates is fulfilled by linear optics; however, the corresponding linear-optical schemes are probabilistic \cite{EPP-linear1,EPP-linear2}. In the third group, cross-Kerr nonlinearities \cite{EPP3,Kerr13,EPP-Kerr1}, matter platforms \cite{NV}, or hyperentanglements \cite{hyperentangled} are employed to overcome the intrinsic probabilistic nature of linear-optical EPPs. There are also several other interesting EPPs, such as measurement-based EPPs \cite{measurement-based-EPPs1,measurement-based-EPPs2}, hyper-parallel EPPs \cite{Ecker2021,hyper-parallel-EPPs26}, multi-copy EPPs \cite{multi-copy-EPPs28,multi-copy-EPPs29,multi-copy-EPPs30}, optimal EPPs \cite{optimal-EPPs33,optimal-EPPs35}, direct EPPs \cite{EPP5}, residual EPPs \cite{residual18,Zhou2025}, and high-dimensional EPPs \cite{high-dimensional-EPPs1,high-dimensional-EPPs2}.

The current EPPs mostly focus on bipartite two-dimensional qubit systems. It is known that multipartite entangled states are the core resources in quantum networks, measurement-based quantum computation, quantum key distribution, dense coding, controlled quantum teleportation, and distributed quantum computation \cite{network,controlled-teleportation,distribute1,distribute2,measurement-based}. Compared with two-dimensional qubit systems, high-dimensional quantum systems (so-called qudits, i.e., $d$-level or $d$-state systems) offer distinct advantages. Specifically, they can significantly enlarge the information capacity \cite{Hu2018,Du1,Du2,Du3}, improve the security in quantum communication \cite{Guo2018,Liu2022}, increase the resilience against environmental noise \cite{Ecker2021}, improve the efficiency and accuracy in quantum tasks \cite{Karacsony2024,d-level55,Zhao,Hong2023}, and strengthen the violation of Bell inequalities \cite{Bell-test1,Bell-test2}. Higher-dimensional entanglement also has the potential to overcome the limitations of qubit entanglement in performing certain QIP tasks, such as quantum stochastic communication \cite{Zhang2025} and complete entanglement analysis with linear optics \cite{overcome1}. To leverage these high-dimensional advantages, various degrees of freedom (DOFs) of single-photon systems have been extensively explored, such as orbital angular momentum \cite{He2022_LSA}, frequency \cite{Kues2017}, time-bin \cite{Islam2017}, and path modes \cite{Wang2018_Science}. Furthermore, high-dimensional Bell states and Greenberger-Horne-Zeilinger (GHZ) states have been experimentally demonstrated across multiple DOFs \cite{HBell,HGHZ}. 
Recently, $d$-level $n$-partite photonic GHZ state has been experimentally generated by using single-photon sources and linear operation \cite{Chin2024npj,Paesani2021,Chin2024}.

In this paper, we present EPP schemes for arbitrary $d$-dimensional $n$-partite GHZ state.  
By using quantum nondemolition (QND) measurement with cross-Kerr nonlinearities, the qudit-flip error and phase-flip error are corrected in two steps. 
We first detail the EPPs for tripartite qutrit-based (i.e., 3-level or 3-state system) GHZ states which have one and two polluted photons, and then extend the program to arbitrary $d$-dimensional $n$-partite GHZ state with $n$ polluted photons scenario.
The proposed schemes have some characteristics. 
Beyond multipartite qubit-based EPP protocols \cite{Kerr13}, our program focuses on arbitrary $n$-partite qudit-based EPPs scenarios. 
The fidelity can be further improved to unity by iterating the EPP process. 
The fidelity thresholds for the proposed EPPs are presented. 
Moreover, our schemes are immune to the number of the polluted photons.

\section{General $d$-dimensional $n$-partite entanglement purification for correcting qudit-flip errors}\label{sec2}

\subsection{Quantum nondemolition detector with cross-Kerr nonlinearity} \label{sec2-1}

Quantum nondemolition detector (QND) with cross-Kerr nonlinearity is a promising tool for implementing some QIP tasks, such as CNOT gate, complete entangled states analysis, EPP, and ECP.
The Hamiltonian for describing the interaction between signal state $|n_s\rangle$ (Fock state, i.e., the signal field contains $n$ photons) and the coherent state $|\alpha_p\rangle$ is given by \cite{Hamiltonian1,Hamiltonian2,Hamiltonian3}
\begin{eqnarray} \label{eq1}
H = \hbar \chi \hat{a}_s^\dagger \hat{a}_s \hat{a}_p^\dagger \hat{a}_p.
\end{eqnarray}
Here $\hat{a}_s^\dagger$ ($\hat{a}_p^\dagger$) and $\hat{a}_s$ ($\hat{a}_p$) are the creation and annihilation operators of the signal (probe) mode $s$ ($p$), respectively. 
$\chi$ is the coupling strength of the nonlinearity, and it is determined by the cross-Kerr material.   
The evolution of the signal-probe system induced by the cross-Kerr nonlinearity is given by
\begin{eqnarray} \label{eq2}
e^{-\text{i}Ht/\hbar} |n\rangle_s|\alpha\rangle_p = |n\rangle_s \left|\alpha e^{-\text{i} n \theta}\right\rangle_p,
\end{eqnarray}
where $\theta = \chi t$ and $t$ is the interaction time. 
The phase shift $n \theta$, picked up by the coherent state $|\alpha\rangle_p$, is directly proportional to the number of the photons in the signal mode $|n\rangle_s$. 
Such a photon-number-dependent phase shift, particularly the $\pi$  phase shift created by a single-photon pulse, has been experimentally realized in Rydberg media \cite{Tiarks2016SciAdv}. 
Next,  utilizing this feature, we construct QND for designing EPP scheme to purify generalized qutrit-based GHZ states.

\begin{figure}[htbp]
    \centering
    \includegraphics[width=0.46\columnwidth]{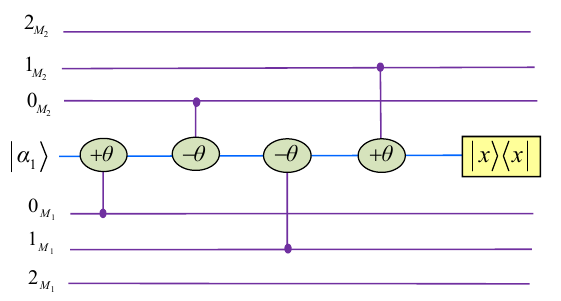} 
    \caption{The principle of the proposed nondestructive quantum nondemolition detector (QND) in three-dimensional systems.
    $|0\rangle_M$, $|1\rangle_M$, and $|2\rangle_M$ ($|0\rangle_{M'}$, $|1\rangle_{M'}$, and $|2\rangle_{M'}$) are the three temporal-DOF of single-photon system $M_1$ ($M_2$). }
    \label{Fig1}
\end{figure}

The principle of QND we designed for high-dimensional EPP is shown in \textbf{Figure~\ref{Fig1}}.  
Here $|0\rangle_M$, $|1\rangle_M$, and $|2\rangle_M$ ($|0\rangle_{M'}$, $|1\rangle_{M'}$, and $|2\rangle_{M'}$) are three temporal-DOF of single-photon system $M$ ($M'$).
The photon-number-dependent phase shift rule induced by the cross-Kerr nonlinearity is shown in \textbf{Table~\ref{Tab0}}.

\begin{table}[htbp]
\centering 
\caption{The photon-number-dependent phase shift rule in the proposed qutrit-based EPPs.}\label{Tab0}
\begin{tabular}{cccccccccc}
\hline \hline  

Input                 & Phase shift      & Input                  & Phase shift            & Input                    & Phase shift \\
\hline
  $|00\rangle_{MM'}$  & 0                & $|10\rangle_{MM'}$     & $-2\theta$             & $|20\rangle_{MM'}$       & $-\theta$  \\

  $|01\rangle_{MM'}$  & $+2\theta$       & $|11\rangle_{MM'}$     & 0                      & $|21\rangle_{MM'}$       & $+\theta$  \\

  $|02\rangle_{MM'}$  & $+\theta$        & $|12\rangle_{MM'}$     & $-\theta$              & $|22\rangle_{MM'}$       & 0  \\

\hline\hline  
\end{tabular}
\end{table}

A set of 27 generalized 3-qutrit GHZ states $\{|\psi_{l,u}^{3,3}\rangle_{A_1B_1C_1}\}$ is given by
\begin{eqnarray} \label{eq3}
 |\psi_{l,u}^{3,3}\rangle_{A_1B_1C_1} = \frac{1}{\sqrt{3}}\sum_{j=0}^{2}{e^{\text{i}\frac{2\pi}{3}lj}}|j\rangle_{A_1}|(j+k_1) \bmod 3\rangle_{B_1}|(j+k_2)\bmod 3\rangle_{C_1},
\end{eqnarray}
where $ l,k_1,k_2\in\{0,1,2\}$ and $u=3k_1+k_2$.
Here and afterwards, in expression $|\psi_{l,u}^{d,n}\rangle_{A_1B_1C_1\cdots N_1}$, 
superscripts $d$ denotes $d$-dimensional and $n$ denotes $n$-partite, 
subscripts $A_1$, $B_1$, $C_1$, $\cdots$, $N_1$ represent the photons $A_1$, $B_1$, $C_1$, $\cdots$, $N_1$ kept by long-distance agents Alice, Bob, Charlie,  $\cdots$, Nick, respectively.


\subsection{EPP for three-qutrit GHZ state with one qutrit-flip error channel}\label{sec2-2}

Suppose that generalized qutrit-flip errors have taken place on photon $C$ in GHZ state $|\psi_{0,0}^{3,3}\rangle_{A_1B_1C_1}$ during transmission, 
and then the desired source state $|\psi_{0,0}^{3,3}\rangle_{A_1B_1C_1}$ will become 
\begin{eqnarray} \label{eq4}
  \begin{split}
  \rho_{A_1B_1C_1} = F_0 |\psi_{0,0}^{3,3}\rangle_{A_1B_1C_1}\langle\psi_{0,0}^{3,3}|
                   + F_1 |\psi_{0,1}^{3,3}\rangle_{A_1B_1C_1}\langle\psi_{0,1}^{3,3}|
                   + F_2 |\psi_{0,2}^{3,3}\rangle_{A_1B_1C_1}\langle\psi_{0,2}^{3,3}|.
  \end{split}
\end{eqnarray}
Based on Eq.~\eqref{eq3}, states $|\psi_{0,0}^{3,3}\rangle$, $|\psi_{0,1}^{3,3}\rangle$, and $|\psi_{0,2}^{3,3}\rangle$ are given by
\begin{eqnarray} \label{eq5}
  \begin{split}
  &|\psi_{0,0}^{3,3}\rangle=\frac{1}{\sqrt{3}}(| 000\rangle +| 111\rangle +| 222\rangle), \\
  &|\psi_{0,1}^{3,3}\rangle=\frac{1}{\sqrt{3}}(| 001\rangle +| 112\rangle +| 220\rangle), \\
  &|\psi_{0,2}^{3,3}\rangle=\frac{1}{\sqrt{3}}(| 002\rangle +| 110\rangle +| 221\rangle).
  \end{split}
\end{eqnarray}
Here $F_0+F_1+F_2=1$. The matrix forms of qutrit-flip errors $X_{3_1}$ and $X_{3_2}$ in the basis $\{|0\rangle, |1\rangle, |2\rangle\}$ are given by
\begin{eqnarray} \label{eq6}
 X_{3_1}=\left(
           \begin{array}{ccc}
             0 & 0 & 1 \\
             1 & 0 & 0 \\
             0 & 1 & 0 \\
          \end{array}
         \right),\quad 
 X_{3_2}=\left(
           \begin{array}{ccc}
            0 & 1 & 0 \\
            0 & 0 & 1 \\
            1 & 0 & 0 \\
           \end{array}
          \right).                 
\end{eqnarray}
Eq. \eqref{eq6} indicates that errors $X_{3_1}$ and $X_{3_2}$ take place with probabilities of $F_1$ and $F_2$,  respectively.

In order to protect state $|\psi_{0,0}^{3,3}\rangle_{ABC}$ from the qutrit-flip channel noise, Alice, Bob, and  Charlie should shared another identical entangled mixed state
\begin{eqnarray} \label{eq7}
  \begin{split}
  \rho_{A_2B_2C_2} =  & F_0 |\psi_{0,0}^{3,3}\rangle_{A_2B_2C_2}\langle\psi_{0,0}^{3,3}| 
                      + F_1 |\psi_{0,1}^{3,3}\rangle_{A_2B_2C_2}\langle\psi_{0,1}^{3,3}|  
                      + F_2 |\psi_{0,2}^{3,3}\rangle_{A_2B_2C_2}\langle\psi_{0,2}^{3,3}|,
  \end{split}
\end{eqnarray} 
where the subscripts $A_2$, $B_2$, and $C_2$ represent the photons processed by Alice, Bob, and Charlie, respectively.
Based on Eqs.~\eqref{eq4} and~\eqref{eq7}, the composite system composed of $A_1$, $B_1$, $C_1$, $A_2$, $B_2$, and $C_2$ can be expressed as
\begin{equation} \label{eq8}
\rho_{A_1B_1C_1} \otimes \rho_{A_2B_2C_2} = \sum_{u,v=0}^{2} F_u F_v \bigl(|\psi_{0,u}^{3,3}\rangle\langle\psi_{0,u}^{3,3}|\bigr)_{A_1B_1C_1} \otimes 
                                                                     \bigl(|\psi_{0,v}^{3,3}\rangle\langle\psi_{0,v}^{3,3}|\bigr)_{A_2B_2C_2}.
\end{equation}

\begin{figure}[htbp]
    \centering
    \includegraphics[width=0.60\textwidth]{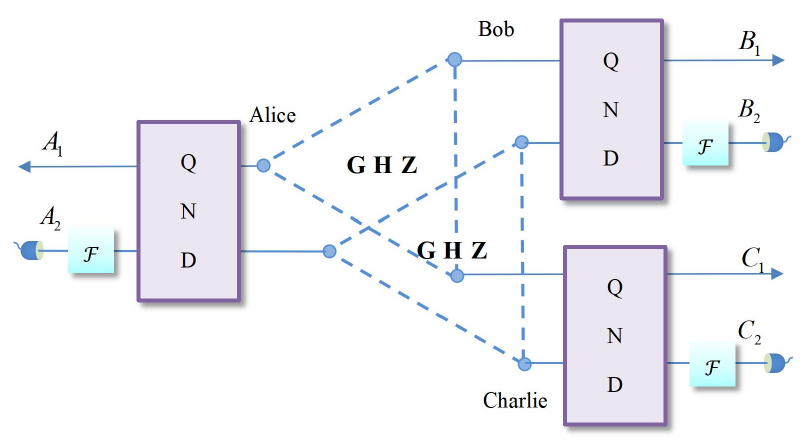}
    \caption{Schematic principle of the entanglement purification protocol for correcting qutrit-flip errors in three-dimensional GHZ states.}
    \label{Fig2}
\end{figure}

The principle of our three-dimensional tripartite EPP protocol to correct qutrit-flip error is shown in \textbf{Figure~ \ref{Fig2}}, and 
Alice, Bob, and Charlie each hold the same setup depicted by Figure~ \ref{Fig1}. 
Let us detail the procedure of our EPP via QNDs, step by step.

In the first step, photon $A_1$ and $A_2$ are injected into the three up spatial modes (i.e., $|0\rangle_{A_1}$, $|1\rangle_{A_1}$, $|2\rangle_{A_1}$) and the down three spatial modes (i.e., $|0\rangle_{A_2}$, $|1\rangle_{A_2}$, $|2\rangle_{A_2}$), respectively.  
So do Bob and Charlie.

In the second step,  after the Kerr nonlinearities, 
Alice, Bob, and Charlie perform $X$-homodyne measurements on coherent states $|\alpha\rangle_A$, $|\alpha\rangle_B$, and $|\alpha\rangle_C$, 
and they only keep the phase shifters $(0_A,0_B,0_C)$. 
Then, the composite systems $\rho_{A_1B_1C_1} \otimes \rho_{A_2B_2C_2}$ will collapse into the following new mixed state 
\begin{eqnarray} \label{eq9}
\begin{split}
 \frac{1}{3}F_0^2 |\phi_{0}^{3,3}\rangle_{A_1B_1C_1A_2B_2C_2}\langle\phi_{0}^{3,3}| 
+\frac{1}{3}F_1^2 |\phi_{1}^{3,3}\rangle_{A_1B_1C_1A_2B_2C_2}\langle\phi_{1}^{3,3}|  
+\frac{1}{3}F_2^2 |\phi_{2}^{3,3}\rangle_{A_1B_1C_1A_2B_2C_2}\langle\phi_{2}^{3,3}|,
\end{split}
\end{eqnarray}
where
\begin{eqnarray} \label{eq10}
\begin{split}
|\phi_0^{3,3}\rangle = \frac{1}{\sqrt{3}}(|000\rangle^{\otimes 2} 
                                  + |111\rangle^{\otimes 2} 
                                  + |222\rangle^{\otimes 2})_{A_1B_1C_1A_2B_2C_2},
\end{split}
\end{eqnarray}
\begin{eqnarray} \label{eq11}
\begin{split}
|\phi_1^{3,3}\rangle = \frac{1}{\sqrt{3}}(|001\rangle^{\otimes 2} 
                                  + |112\rangle^{\otimes 2} 
                                  + |220\rangle^{\otimes 2})_{A_1B_1C_1A_2B_2C_2},
\end{split}
\end{eqnarray}
\begin{eqnarray} \label{eq12}
\begin{split}
|\phi_2^{3,3}\rangle = \frac{1}{\sqrt{3}}(|002\rangle^{\otimes 2} 
                                  + |110\rangle^{\otimes 2} 
                                  + |221\rangle^{\otimes 2})_{A_1B_1C_1A_2B_2C_2}.
\end{split}
\end{eqnarray}
Note that, all the cross combinations $|\psi_{0,u}^{3,3}\rangle_{A_1B_1C_1}\otimes|\psi_{0,v}^{3,3}\rangle_{A_2B_2C_2},(u\neq v)$ are wiped out.

In the third step, qutrit-based Fourier transformations $\mathcal{F}_3$s are performed on photon $A_2$, $B_2$, and $C_2$ which are used to accomplish the following transformations
\begin{eqnarray}    \label{eq13}
 \begin{split}
  & |0\rangle \xrightarrow{\mathcal{F}_3} \frac{1}{\sqrt{3}}(|0\rangle + |1\rangle + |2\rangle),\\
  & |1\rangle \xrightarrow{\mathcal{F}_3} \frac{1}{\sqrt{3}}(|0\rangle + e^{\text{i}\frac{2\pi}{3}}|1\rangle +e^{\text{i}\frac{4\pi}{3}}|2\rangle) ,\\
  & |2\rangle \xrightarrow{\mathcal{F}_3}  \frac{1}{\sqrt{3}}(|0\rangle + e^{\text{i}\frac{4\pi}{3}}|1\rangle + e^{\text{i}\frac{2\pi}{3}}|2\rangle).
\end{split}
\end{eqnarray}

In the fourth step, based on the measurement results, some classical feed-forward operations (see \textbf{Table \ref{Tab1}}) are performed on the photons $A_1$, $B_1$, and $C_1$ to obtain the desired state $|\psi_{0,0}^{3,3}\rangle_{A_1B_1C_1}$ with an improved fidelity of $\frac{F_0^2}{F_0^2 + F_1^2 + F_2^2}>F_0$ with $F_0>\frac{1}{3}$. The proof is presented in Appendix \ref{secA}.

\begin{table}[htbp]
  \centering 
  \caption{The correspondence between the measurement results of photon pairs ($A_2$, $B_2$,  $C_2$), the output states of photon pairs ($A_1$, $B_1$, $C_1$), 
           and the classical feed-forward operations for the state $|\phi_0^{3,3}\rangle_{A_1B_1C_1A_2B_2C_2}\langle\phi_0^{3,3}|$ given in Eq. \eqref{eq9}. 
           Here single-qutrit operations 
           $I_3 = |0\rangle\langle0| + |1\rangle\langle1| + |2\rangle\langle2|$, 
           $Z_{3_1} = |0\rangle\langle0| + e^{\text{i}\frac{2\pi}{3}}|1\rangle\langle1| + e^{\text{i}\frac{4\pi}{3}} |2\rangle\langle2|$, and 
           $Z_{3_2} = |0\rangle\langle0| + e^{\text{i}\frac{4\pi}{3}}|1\rangle\langle1| + e^{\text{i}\frac{2\pi}{3}} |2\rangle\langle2|$.}      
  \label{Tab1}
  \begin{tabular}{ccccccc}
  
  \hline  \hline 

  \multirow{2}*{Measurement}  &   \multirow{2}*{Outputs}           & \multicolumn {3}{c}{Feed-forward} \\  \cline{3-5}
  
                              &                                    &    $A_1$      &  $B_1$   &   $C_1$  \\
  \hline
                              
  $|0 0 0\rangle_{A_2B_2C_2}$ &  $|\psi_{0,0}^{3,3}\rangle_{A_1B_1C_1}$  &    $I_3$      &   $I_3$   &  $I_3$  \\
  $|0 0 1\rangle_{A_2B_2C_2}$ &  $|\psi_{1,0}^{3,3}\rangle_{A_1B_1C_1}$  &    $Z_{3_2}$  &   $I_3$   &  $I_3$  \\
  $|0 0 2\rangle_{A_2B_2C_2}$ &  $|\psi_{2,0}^{3,3}\rangle_{A_1B_1C_1}$  &    $Z_{3_1}$  &   $I_3$   &  $I_3$  \\
  $|0 1 0\rangle_{A_2B_2C_2}$ &  $|\psi_{1,0}^{3,3}\rangle_{A_1B_1C_1}$  &    $Z_{3_2}$  &   $I_3$   &  $I_3$  \\
  $|0 1 1\rangle_{A_2B_2C_2}$ &  $|\psi_{2,0}^{3,3}\rangle_{A_1B_1C_1}$  &    $Z_{3_1}$  &   $I_3$   &  $I_3$  \\
  $|0 1 2\rangle_{A_2B_2C_2}$ &  $|\psi_{0,0}^{3,3}\rangle_{A_1B_1C_1}$  &    $I_3$      &   $I_3$   &  $I_3$  \\
  $|0 2 0\rangle_{A_2B_2C_2}$ &  $|\psi_{2,0}^{3,3}\rangle_{A_1B_1C_1}$  &    $Z_{3_1}$  &   $I_3$   &  $I_3$  \\
  $|0 2 1\rangle_{A_2B_2C_2}$ &  $|\psi_{0,0}^{3,3}\rangle_{A_1B_1C_1}$  &    $I_3$      &   $I_3$   &  $I_3$  \\
  $|0 2 2\rangle_{A_2B_2C_2}$ &  $|\psi_{1,0}^{3,3}\rangle_{A_1B_1C_1}$  &    $Z_{3_1}$  &   $I_3$   &  $I_3$  \\

  $|1 0 0\rangle_{A_2B_2C_2}$ &  $|\psi_{1,0}^{3,3}\rangle_{A_1B_1C_1}$  &    $Z_{3_2}$  &   $I_3$   &  $I_3$  \\
  $|1 0 1\rangle_{A_2B_2C_2}$ &  $|\psi_{2,0}^{3,3}\rangle_{A_1B_1C_1}$  &    $Z_{3_1}$  &   $I_3$   &  $I_3$  \\
  $|1 0 2\rangle_{A_2B_2C_2}$ &  $|\psi_{0,0}^{3,3}\rangle_{A_1B_1C_1}$  &    $I_3$      &   $I_3$   &  $I_3$  \\
  $|1 1 0\rangle_{A_2B_2C_2}$ &  $|\psi_{2,0}^{3,3}\rangle_{A_1B_1C_1}$  &    $Z_{3_2}$  &   $I_3$   &  $I_3$  \\
  $|1 1 1\rangle_{A_2B_2C_2}$ &  $|\psi_{0,0}^{3,3}\rangle_{A_1B_1C_1}$  &    $I_3$      &   $I_3$   &  $I_3$  \\
  $|1 1 2\rangle_{A_2B_2C_2}$ &  $|\psi_{1,0}^{3,3}\rangle_{A_1B_1C_1}$  &    $Z_{3_2}$  &   $I_3$   &  $I_3$  \\
  $|1 2 0\rangle_{A_2B_2C_2}$ &  $|\psi_{0,0}^{3,3}\rangle_{A_1B_1C_1}$  &    $I_3$      &   $I_3$   &  $I_3$  \\
  $|1 2 1\rangle_{A_2B_2C_2}$ &  $|\psi_{1,0}^{3,3}\rangle_{A_1B_1C_1}$  &    $Z_{3_2}$  &   $I_3$   &  $I_3$  \\
  $|1 2 2\rangle_{A_2B_2C_2}$ &  $|\psi_{2,0}^{3,3}\rangle_{A_1B_1C_1}$  &    $Z_{3_1}$  &   $I_3$   &  $I_3$  \\
  
  $|2 0 0\rangle_{A_2B_2C_2}$ &  $|\psi_{2,0}^{3,3}\rangle_{A_1B_1C_1}$  &    $Z_{3_1}$  &   $I_3$   &  $I_3$  \\
  $|2 0 1\rangle_{A_2B_2C_2}$ &  $|\psi_{0,0}^{3,3}\rangle_{A_1B_1C_1}$  &    $I_3$      &   $I_3$   &  $I_3$  \\
  $|2 0 2\rangle_{A_2B_2C_2}$ &  $|\psi_{1,0}^{3,3}\rangle_{A_1B_1C_1}$  &    $Z_{3_2}$  &   $I_3$   &  $I_3$  \\
  $|2 1 0\rangle_{A_2B_2C_2}$ &  $|\psi_{0,0}^{3,3}\rangle_{A_1B_1C_1}$  &    $I_3$      &   $I_3$   &  $I_3$  \\
  $|2 1 1\rangle_{A_2B_2C_2}$ &  $|\psi_{1,0}^{3,3}\rangle_{A_1B_1C_1}$  &    $Z_{3_2}$  &   $I_3$   &  $I_3$  \\
  $|2 1 2\rangle_{A_2B_2C_2}$ &  $|\psi_{2,0}^{3,3}\rangle_{A_1B_1C_1}$  &    $Z_{3_1}$  &   $I_3$   &  $I_3$  \\
  $|2 2 0\rangle_{A_2B_2C_2}$ &  $|\psi_{1,0}^{3,3}\rangle_{A_1B_1C_1}$  &    $Z_{3_2}$  &   $I_3$   &  $I_3$  \\
  $|2 2 1\rangle_{A_2B_2C_2}$ &  $|\psi_{2,0}^{3,3}\rangle_{A_1B_1C_1}$  &    $Z_{3_1}$  &   $I_3$   &  $I_3$  \\
  $|2 2 2\rangle_{A_2B_2C_2}$ &  $|\psi_{0,0}^{3,3}\rangle_{A_1B_1C_1}$  &    $I_3$      &   $I_3$   &  $I_3$  \\
\hline \hline  
  \end{tabular}
\end{table}


\subsection{EPP for $n$-qudit GHZ state with one qudit-flip error channel}\label{sec2-3}

For the arbitrary  $d$-dimensional $n$-partite GHZ state, a complete set of generalized GHZ basis states can be compactly written as
\begin{eqnarray} \label{eq14}
|\psi_{l,u}^{d,n}\rangle_{A_1B_1\dots N_1} = \frac{1}{\sqrt{d}} 
  \sum_{j=0}^{d-1} e^{\text{i}\frac{2\pi}{d}lj} |j\rangle_{A_1} \otimes \big(\bigotimes_{r=1}^{n-1} |(j + k_r) \bmod d\rangle \big)_{B_1 C_1 \cdots N_1},
\end{eqnarray}
%
%
where $l\in\{0,1,\ldots,(d-1)\}$,
      $k_1,k_2,\cdots, k_{n-1}\in\{0,1,\ldots,(d-1)\}$, 
      $u\in\{0,1,\ldots, (d^{n-1}-1)\}$, and the base-10 number $u$ is given by
\begin{eqnarray} \label{eq15}
u=\sum_{i=1}^{n-1} k_i \cdot d^{(n-1-i)} = k_1 \cdot d^{n-2} + k_2 \cdot d^{n-3} + \cdots + k_{n-1} d^{0}.
\end{eqnarray}

Our program can be extended to distill the arbitrary $d$-dimensional $n$-partite  GHZ state $|\psi_{0,0}^{d,n}\rangle_{A_1B_1\cdots N_1}$ shown with one qudit-flip polluted photon. 
The qudit-flip error $X_{d_i}$ acting on the last particle can be expressed as
\begin{eqnarray} \label{eq16}
  \begin{split}
   X_{d_i} = \sum_{j=0}^{d-1} | (j + i) \bmod d\rangle  \langle j |, 
  \end{split}
\end{eqnarray}
where $i=0,1,\cdots, (d-1)$. These errors make $|\psi_{0,0}^{d,n}\rangle_{A_1B_1\cdots N_1}$ become
\begin{equation}\label{eq17}
\dot{\rho}_{A_1B_1\cdots N_1} = \sum_{u=0}^{d-1} \dot{F}_u |\psi_{0,u}^{d,n}\rangle_{A_1B_1\cdots N_1} \langle \psi_{0,u}^{d,n}|.
\end{equation}
Based on Eq. \eqref{eq14},  state $|\psi_{0, u}^{d,n}\rangle_{A_1 \cdots N_1}$ can be reedited as
\begin{eqnarray} \label{eq18}
|\psi_{0, u}^{d,n}\rangle_{A_1 \cdots N_1} = \frac{1}{\sqrt{d}} \sum_{j=0}^{d-1} (|j\rangle^{\otimes (n-1)} \otimes |(j+u) \bmod d\rangle)_{A_1 B_1 \cdots N_1}.
\end{eqnarray}
%

\begin{figure}[htbp]
    \centering
    \includegraphics[width=0.8\textwidth]{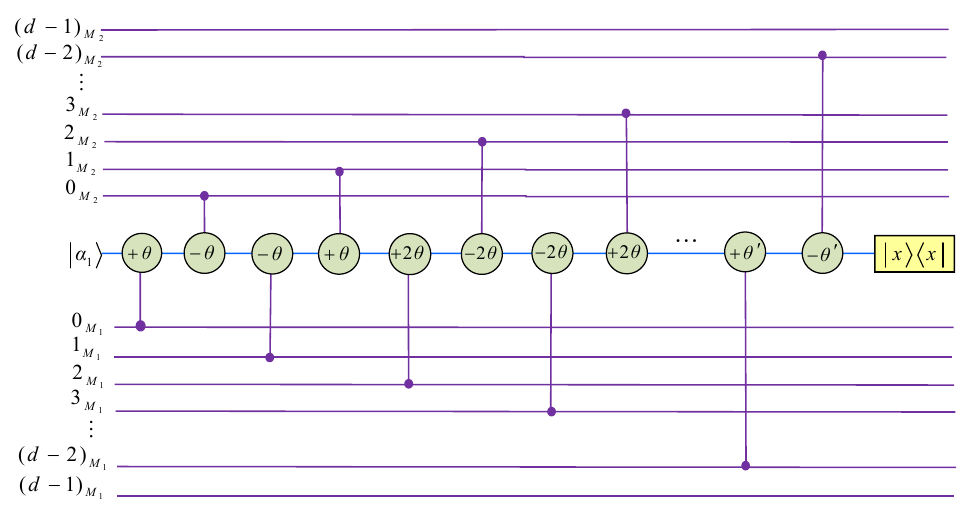}
    \caption{The principle of the proposed QND in qudit systems.
             Phase shifts $+\theta$, $-\theta$,  $+2\theta$, $-2\theta$, $+3\theta$, $-3\theta$,  $+4\theta$, $-4\theta$, $\cdots$, $\theta'$  are introduced to the states 
                          $0_{M_1}$, $1_{M_1}$,  $2_{M_1}$,  $3_{M_1}$,  $4_{M_1}$,  $5_{M_1}$,  $6_{M_1}$,   $7_{M_1}$,  $\cdots$, $(d-2)_{M_1}$, respectively.
             Phase shifts $-\theta$, $+\theta$,  $-2\theta$, $+2\theta$, $-3\theta$, $+3\theta$,  $-4\theta$, $+4\theta$, $\cdots$, $-\theta'$ are introduced to the states 
                          $0_{M_2}$, $1_{M_2}$,  $2_{M2}$,   $3 _{M2}$,  $4_{M_2}$,  $5_{M_2}$,  $6_{M_2}$,   $7_{M_2}$,  $\cdots$, $(d-2)_{M_2}$, respectively.
             Here $\theta'=(-1)^n\left\lceil\frac{n+1}{2}\right\rceil \theta$ with $n=0,1,\cdots, (d-2)$.} 
    \label{Fig3}
\end{figure}

For the qudit-flip error $X_{d_i}$, 
we first should extend the qutrit-based QND equipments to the qudit-based scenario, see \textbf{Figure~ \ref{Fig3}}, 
and then we follow the similar arrangement as that made in Sec. \ref{sec2-2}.

In detail, a noisy copy $\dot{\rho}_{A_2B_2\cdots N_2}$  with the same form as Eq. \eqref{eq17} is introduced, i.e.,
\begin{equation}\label{eq19}
\dot{\rho}_{A_2B_2\cdots N_2} = \sum_{v=0}^{d-1} \dot{F}_{v} |\psi_{0,v}^{d,n}\rangle_{A_2B_2\cdots N_2} \langle \psi_{0,v}^{d,n}|.
\end{equation}
Based on Eq. \eqref{eq14}, state $|\psi_{0, v}^{d,n}\rangle_{A_1 \cdots N_1}$ can be reedited as
\begin{eqnarray} \label{eq20}
|\psi_{0, v}^{d,n}\rangle_{A_1 \cdots N_1} = \frac{1}{\sqrt{d}} \sum_{j=0}^{d-1}(|j\rangle^{\otimes (n-1)} \otimes |(j+v) \bmod d\rangle)_{A_1 B_1 \cdots N_1}.
\end{eqnarray}
%

After the qudit-based QND equipments performed on photon pairs ($A_1$, $A_2$), ($B_1$, $B_2$), $\dots$, ($N_1$, $N_2$), 
parties Alice, Bob, $\cdots$, Nick choose the phases $(0_A, 0_B, 0_C,\cdots, 0_N)$.
Then the composite system $\dot{\rho}_{A_1B_1\cdots N_1} \otimes \dot{\rho}_{A_2B_2\cdots N_2} $ will be projected into the following new mixed state
\begin{eqnarray} \label{eq21}
\begin{split}
\sum_{u=0}^{d-1} \frac{1}{d}{\dot{F}_{u}}^2 |\dot{\phi}_{u}^{d,n}\rangle_{A_1B_1\cdots N_1A_2B_2\cdots N_2} \langle \dot{\phi}_{u}^{d,n}| ,
\end{split}
\end{eqnarray}
where
\begin{eqnarray} \label{eq22}
|\dot{\phi}_{u}^{d,n}\rangle = \frac{1}{\sqrt d} \sum_{j=0}^{d-1} (|j\rangle^{\otimes (n-1)} \otimes |(j+u)\bmod d\rangle)^{\otimes 2}_{A_1B_1\cdots N_1A_2B_2\cdots N_2}.
\end{eqnarray}
That is, after QND,  all the cross combinations $|\psi_{0,u}^{d,n}\rangle_{A_1B_1C_1}\otimes|\psi_{0,v}^{d,n}\rangle_{A_2B_2C_2}, (u \neq v)$ are wiped out.

Subsequently,  qudit-based Fourier transformations $\mathcal{F}_{d}$s are performed on photon $A_2$, $B_2$, $\dots$, $N_2$. Here $\mathcal{F}_{d}$ completes the transformation
\begin{eqnarray} \label{eq23}
\begin{split}
 \mathcal{F}_{d} |\lambda\rangle = \frac{1}{\sqrt{d}} \sum_{k=0}^{d-1} \omega_{d}^{k\lambda} |k\rangle, 
\end{split}
\end{eqnarray}
where $\omega_d = e^{\text{i}\frac{2\pi}{d}}$, and $\lambda=0,1, \cdots, (d-1)$.

Nextly, photons $A_2$, $B_2$, $\dots$, $N_2$ are detected in the basis $\{| 0\rangle, | 1\rangle, \cdots, | d-1\rangle\}$.

Lastly, based on above measurement results, some suitably classical feed-forward single-qudit operations are performed on photon $A_1$, $B_1$, $\dots$, $N_1$
to obtain the desired output state $|\psi_{0,0}^{d,n}\rangle_{A_1B_1\cdots N_1}$ with an enhanced fidelity of 
 $\frac{\dot{F}_{0}^2}{\sum_{u=0}^{d-1} \dot{F}_{u}^2} > \dot{F}_{0}$ with $\dot{F}_{0}>\frac{1}{d}$.  The detailed proof process can be found in Appendix  \ref{secB}.

\subsection{EPP for $n$-qudit GHZ state with $n$ nontrivial qudit-flip errors channels}\label{sec2-4}

Suppose that all the photons in the $d$-dimensional $n$-partite GHZ state $|\psi_{0,0}^{d,n}\rangle_{A_1B_1\cdots N_1}$suffer from arbitrary qudit-flip errors. After the transmission of photons, $|\psi_{0,0}^{d,n}\rangle_{A_1B_1\cdots N_1}$ becomes
\begin{eqnarray} \label{eq24}
\ddot{\rho}_{A_1B_1\cdots N_1} = \sum_{u=0}^{d^{n-1}-1} \ddot{F}_{u} |\psi_{0,u}^{d,n}\rangle \langle \psi_{0,u}^{d,n} |.
\end{eqnarray}
Here $|\psi_{0,u}^{d,n}\rangle$ is given by Eq. \eqref{eq14}.
To correct these errors, a noisy copy $\ddot{\rho}_{A_2B_2\cdots N_2}$ with the same form as Eq. \eqref{eq24} is required.

We use the same arrangement  as that made in Sec. \ref{sec2-3}.
After the QND equipments shown in Figure~\ref{Fig1} are performed on photon pairs $(A_1,A_2)$, $(B_1,B_2)$, $\cdots$, $(N_1,N_2)$, 
parties Alice, Bob, $\cdots$, Nick only keep the same phase outcomes $(0_A,0_B,\cdots,0_N)$, 
and then the composite system $\ddot{\rho}_{A_1B_1\cdots N_1}\otimes \ddot{\rho}_{A_2B_2\cdots N_2}$ will be projected into the following new mixed state 
\begin{eqnarray} \label{eq25}
\begin{split}
\sum_{u=0}^{d^{n-1}-1} \frac{1}{d}{\ddot{F}_u}^2 |\ddot{\phi}_u^{d,n}\rangle_{A_1B_1\cdots N_1A_2B_2\cdots N_2} \langle \ddot{\phi}_u^{d,n}| ,
\end{split}
\end{eqnarray}
where
\begin{eqnarray} \label{eq26}
|\ddot{\phi}_{u}^{d,n}\rangle = \frac{1}{\sqrt d} \sum_{j=0}^{d-1} \big( |j\rangle \otimes (\bigotimes_{r=1}^{n-1} |(j + k_r) \bmod d\rangle) \big)^{\otimes 2}_{A_1 \cdots N_1 A_2 \cdots N_2},
\end{eqnarray}
with $k_1,k_2,\cdots, k_{n-1}\in\{0,1,\ldots,(d-1)\}$.

Fourier transformations $\mathcal{F}_d$ are then performed on photons $A_2$, $B_2$, $\cdots$, $N_2$, respectively. Subsequently, photons $A_2$, $B_2$, $\cdots$, $N_2$ are detected in the basis $\{|0\rangle,|1\rangle,\cdots,|d-1\rangle\}$. After the detected-dependent classical single-qudit feed-forward operations are performed on photons $A_1$, $B_1$, $\cdots$, $N_1$, the desired output $|\psi_{0,0}^{d,n}\rangle_{A_1B_1\cdots N_1}$ can be obtained with an improved fidelity
$\frac{\ddot{F}_0^2}{\sum_{u=0}^{d^{n-1}-1}\ddot{F}_u^2}$. 
Note that $\frac{\ddot{F}_0^2}{\sum_{u=0}^{d^{n-1}-1}\ddot{F}_{u}^2}> \ddot{F}_0$ if and only if $\ddot{F}_0 > \frac{1}{d^{n-1}}$, and the detailed proof process is presented in Appendix \ref{secC}.


\section{General $d$-dimensional $n$-partite entanglement purification for correcting phase-flip errors}\label{sec3}

\subsection{EPP for three-qutrit GHZ state with phase-flip error channels }\label{sec3-1}

Consider photons involved in the state $|\psi_{0,0}^{3,3}\rangle_{A_1B_1C_1}$ suffer from the qutrit-based phase-flip errors $Z_{3_1}$ and $Z_{3_2}$. 
Here
\begin{eqnarray} \label{eq27}
  Z_{3_1}=\left(
              \begin{array}{ccc}
              1 &              0             &            0               \\
              0 & e^{\text{i}\frac{2\pi}{3}} &            0               \\
              0 &              0             & e^{\text{i}\frac{4\pi}{3}} \\
              \end{array}
          \right),\quad 
  Z_{3_2}=\left(
               \begin{array}{ccc}
                1 &              0             &            0               \\
                0 & e^{\text{i}\frac{4\pi}{3}} &            0               \\
                0 &              0             & e^{\text{i}\frac{2\pi}{3}} \\
                \end{array}
           \right).                     
\end{eqnarray}
After the transmission of three photons $A_1$, $B_1$, and $C_1$, the source GHZ state $|\psi_{0,0}^{3,3}\rangle_{A_1B_1C_1}$ will become a mixed state 
\begin{eqnarray} \label{eq28}
  \tilde{\rho}_{A_1B_1C_1} = & \tilde{F}_{0}|\psi_{0,0}^{3,3}\rangle_{A_1B_1C_1}\langle\psi_{0,0}^{3,3}| 
                             + \tilde{F}_{1}|\psi_{1,0}^{3,3}\rangle_{A_1B_1C_1}\langle\psi_{1,0}^{3,3}| 
                             + \tilde{F}_{2}|\psi_{2,0}^{3,3}\rangle_{A_1B_1C_1}\langle\psi_{2,0}^{3,3}|.
\end{eqnarray}
Based on Eq.~\eqref{eq3}, states $|\psi_{0,0}^{3,3}\rangle$, $|\psi_{1,0}^{3,3}\rangle$, and $|\psi_{2,0}^{3,3}\rangle$ are given by
\begin{eqnarray} \label{eq29}
\begin{split}
  &|\psi_{0,0}^{3,3}\rangle=\frac{1}{\sqrt{3}}(|000\rangle                           +|111\rangle                           +|222\rangle),\\ 
  &|\psi_{1,0}^{3,3}\rangle=\frac{1}{\sqrt{3}}(|000\rangle +e^{\text{i}\frac{2\pi}{3}}|111\rangle +e^{\text{i}\frac{4\pi}{3}}|222\rangle),\\ 
  &|\psi_{2,0}^{3,3}\rangle=\frac{1}{\sqrt{3}}(|000\rangle +e^{\text{i}\frac{4\pi}{3}}|111\rangle +e^{\text{i}\frac{2\pi}{3}}|222\rangle).
\end{split}
\end{eqnarray}

\begin{figure}[htbp]
    \centering
    \includegraphics[width=0.65\textwidth]{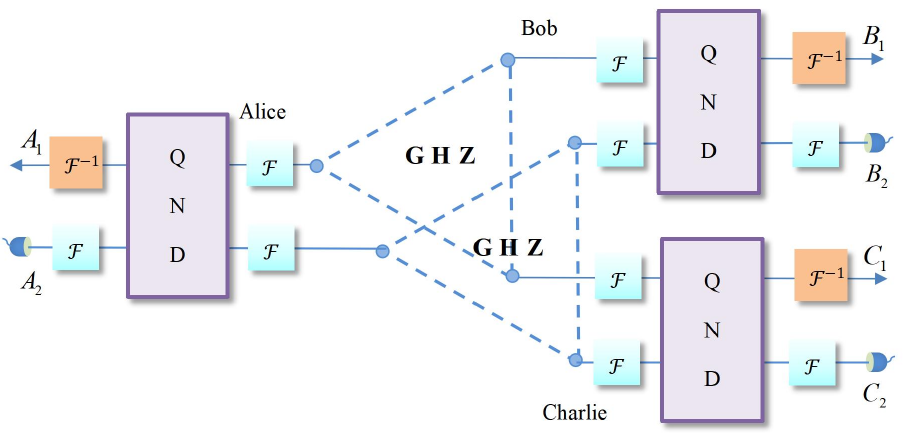}
    \caption{Schematic of the entanglement purification protocol for rectifying phase-flip errors in three-dimensional GHZ states.}
  
    \label{Fig4}
\end{figure}
 
\begin{table}[htbp]
  \centering 
  \caption{The correspondence between the measurement results of photon pairs ($A_2$, $B_2$,  $C_2$), the output states of photon pairs ($A_1$, $B_1$, $C_1$), 
           and the classical feed-forward operations for the state $|\tilde{\phi}_0^{3,3}\rangle_{A_1B_1C_1A_2B_2C_2}\langle\tilde{\phi}_0^{3,3}|$ given in Eq. \eqref{eq33}. 
           Here single-qutrit operation $X_{3_1} = |1\rangle\langle0| + |2\rangle\langle1| + |0\rangle\langle2|$ and 
                                        $X_{3_2}= |2\rangle\langle0| + |0\rangle\langle1| + |1\rangle\langle2|$.}
  \label{Tab2}
  \begin{tabular}{ccccccc}
  
  \hline 

  \multirow{2}*{Measurement}   &   \multirow{2}*{Outputs}          & \multicolumn {3}{c}{Feed-forward} \\  \cline{3-5}
  
                               &                                    &  $A_1$      &  $B_1$     &   $C_1$  \\
  \hline
                              
  $|0 0 0\rangle_{A_2B_2C_2}$  &  $|\psi_{0,0}^{3,3}\rangle_{A_1B_1C_1}$  &  $I_3$      &  $I_3$      &  $I_3$  \\
  $|0 0 1\rangle_{A_2B_2C_2}$  &  $|\psi_{0,1}^{3,3}\rangle_{A_1B_1C_1}$  &  $I_3$      &  $I_3$      &  $X_{3_2}$  \\
  $|0 0 2\rangle_{A_2B_2C_2}$  &  $|\psi_{0,2}^{3,3}\rangle_{A_1B_1C_1}$  &  $I_3$      &  $I_3$      &  $X_{3_1}$  \\
  $|0 1 0\rangle_{A_2B_2C_2}$  &  $|\psi_{0,3}^{3,3}\rangle_{A_1B_1C_1}$  &  $I_3$      &  $X_{3_2}$  &  $I_3$  \\
  $|0 1 1\rangle_{A_2B_2C_2}$  &  $|\psi_{0,4}^{3,3}\rangle_{A_1B_1C_1}$  &  $X_{3_1}$  &  $I_3$      &  $I_3$ \\
  $|0 1 2\rangle_{A_2B_2C_2}$  &  $|\psi_{0,5}^{3,3}\rangle_{A_1B_1C_1}$  &  $I_3$      &  $X_{3_2}$  &  $X_{3_1}$  \\
  $|0 2 0\rangle_{A_2B_2C_2}$  &  $|\psi_{0,6}^{3,3}\rangle_{A_1B_1C_1}$  &  $I_3$      &  $X_{3_1}$  &  $I_3$  \\
  $|0 2 1\rangle_{A_2B_2C_2}$  &  $|\psi_{0,7}^{3,3}\rangle_{A_1B_1C_1}$  &  $I_3$      &  $X_{3_1}$  &  $X_{3_2}$  \\
  $|0 2 2\rangle_{A_2B_2C_2}$  &  $|\psi_{0,8}^{3,3}\rangle_{A_1B_1C_1}$  &  $X_{3_1}$  &  $I_3$      &  $I_3$  \\

  $|1 0 0\rangle_{A_2B_2C_2}$  &  $|\psi_{0,8}^{3,3}\rangle_{A_1B_1C_1}$  &  $X_{3_1}$  &   $I_3$     & $I_3$  \\
  $|1 0 1\rangle_{A_2B_2C_2}$  &  $|\psi_{0,6}^{3,3}\rangle_{A_1B_1C_1}$  &  $I_3$      &   $X_{3_1}$ & $I_3$  \\
  $|1 0 2\rangle_{A_2B_2C_2}$  &  $|\psi_{0,7}^{3,3}\rangle_{A_1B_1C_1}$  &  $I_3$      &   $X_{3_1}$ & $X_{3_2}$  \\
  $|1 1 0\rangle_{A_2B_2C_2}$  &  $|\psi_{0,2}^{3,3}\rangle_{A_1B_1C_1}$  &  $I_3$      &   $I_3$     & $X_{3_1}$  \\
  $|1 1 1\rangle_{A_2B_2C_2}$  &  $|\psi_{0,0}^{3,3}\rangle_{A_1B_1C_1}$  &  $I_3$      &   $I_3$     & $I_3$  \\
  $|1 1 2\rangle_{A_2B_2C_2}$  &  $|\psi_{0,1}^{3,3}\rangle_{A_1B_1C_1}$  &  $I_3$      &   $I_3$     & $X_{3_2}$  \\
  $|1 2 0\rangle_{A_2B_2C_2}$  &  $|\psi_{0,5}^{3,3}\rangle_{A_1B_1C_1}$  &  $I_3$      &   $X_{3_2}$ & $X_{3_1}$  \\
  $|1 2 1\rangle_{A_2B_2C_2}$  &  $|\psi_{0,3}^{3,3}\rangle_{A_1B_1C_1}$  &  $I_3$      &   $X_{3_2}$ & $I_3$  \\
  $|1 2 2\rangle_{A_2B_2C_2}$  &  $|\psi_{0,4}^{3,3}\rangle_{A_1B_1C_1}$  &  $X_{3_1}$  &   $I_3$     & $I_3$  \\
  
  $|2 0 0\rangle_{A_2B_2C_2}$  &  $|\psi_{0,4}^{3,3}\rangle_{A_1B_1C_1}$  &  $X_{3_1}$  &   $I_3$     & $I_3$  \\
  $|2 0 1\rangle_{A_2B_2C_2}$  &  $|\psi_{0,5}^{3,3}\rangle_{A_1B_1C_1}$  &  $I_3$      &   $X_{3_2}$ & $X_{3_1}$  \\
  $|2 0 2\rangle_{A_2B_2C_2}$  &  $|\psi_{0,3}^{3,3}\rangle_{A_1B_1C_1}$  &  $I_3$      &   $X_{3_2}$ & $I_3$  \\
  $|2 1 0\rangle_{A_2B_2C_2}$  &  $|\psi_{0,7}^{3,3}\rangle_{A_1B_1C_1}$  &  $I_3$      &   $X_{3_1}$ & $X_{3_2}$ \\
  $|2 1 1\rangle_{A_2B_2C_2}$  &  $|\psi_{0,8}^{3,3}\rangle_{A_1B_1C_1}$  &  $X_{3_1}$  &   $I_3$     & $I_3$  \\
  $|2 1 2\rangle_{A_2B_2C_2}$  &  $|\psi_{0,6}^{3,3}\rangle_{A_1B_1C_1}$  &  $I_3$      &   $X_{3_1}$ & $I_3$ \\
  $|2 2 0\rangle_{A_2B_2C_2}$  &  $|\psi_{0,1}^{3,3}\rangle_{A_1B_1C_1}$  &  $I_3$      &   $I_3$     & $X_{3_2}$ \\
  $|2 2 1\rangle_{A_2B_2C_2}$  &  $|\psi_{0,2}^{3,3}\rangle_{A_1B_1C_1}$  &  $I_3$      &   $I_3$     & $X_{3_1}$  \\
  $|2 2 2\rangle_{A_2B_2C_2}$  &  $|\psi_{0,0}^{3,3}\rangle_{A_1B_1C_1}$  &  $I_3$      &   $I_3$     & $I_3$  \\
  \hline  
  \end{tabular}
\end{table}

A copy $\tilde{\rho}_{A_2B_2C_2}$ with the same form as Eq. \eqref{eq28} is required to correct the phase-flip errors.
Based on Eq. \eqref{eq3}, we can see that the QND equipment shown in Figure~ \ref{Fig1} can not be used to correct the phase-flip errors directly. 
However, it is notable that after Fourier transformations $\mathcal{F}_3$s are performed on each photon, states $|\psi_{0,0}^{3,3}\rangle$, $|\psi_{1,0}^{3,3}\rangle$, $|\psi_{2,0}^{3,3}\rangle$ are converted into the following ones 
\begin{eqnarray} \label{eq30}
  |\psi_{0,0}^{3,3}\rangle \xrightarrow{\mathcal{F}_3^{\otimes 3}} \frac{1}{\sqrt{3} }(|\psi_{0,0}^{3,3}\rangle + |\psi_{0,5}^{3,3}\rangle + |\psi_{0,7}^{3,3}\rangle),
\end{eqnarray}
\begin{eqnarray} \label{eq31}
  |\psi_{1,0}^{3,3}\rangle \xrightarrow{\mathcal{F}_3^{\otimes 3}} \frac{1}{\sqrt{3} }(|\psi_{0,6}^{3,3}\rangle + |\psi_{0,4}^{3,3}\rangle + |\psi_{0,2}^{3,3}\rangle),
\end{eqnarray}
\begin{eqnarray} \label{eq32}
  |\psi_{2,0}^{3,3}\rangle \xrightarrow{\mathcal{F}_3^{\otimes 3}} \frac{1}{\sqrt{3} }(|\psi_{0,8}^{3,3}\rangle + |\psi_{0,1}^{3,3}\rangle + |\psi_{0,3}^{3,3}\rangle).
\end{eqnarray}
Therefore, as shown in \textbf{Figure~ \ref{Fig4}}, the QND equipment shown in Figure~ \ref{Fig1} supplemented with Fourier transformations are sufficient to correct phase-flip errors.
More in detail, after $\mathcal{F}_3$s and QNDs are performed in succession, and Alice, Bob, and Charlie only choose the phase shifters ($0_A$, $0_B$, $0_C$). 
Then the composite system will project into a new  mixed state 
\begin{eqnarray} \label{eq33}
\begin{split}
 \frac{1}{3}\tilde{F}_{0}^2 |\tilde{\phi}_{0}^{3,3}\rangle_{A_1B_1C_1A_2B_2C_2}\langle\tilde{\phi}_{0}^{3,3}| 
+\frac{1}{3}\tilde{F}_{1}^2 |\tilde{\phi}_{1}^{3,3}\rangle_{A_1B_1C_1A_2B_2C_2}\langle\tilde{\phi}_{1}^{3,3}|  
+\frac{1}{3}\tilde{F}_{2}^2 |\tilde{\phi}_{2}^{3,3}\rangle_{A_1B_1C_1A_2B_2C_2}\langle\tilde{\phi}_{2}^{3,3}|,
\end{split}
\end{eqnarray}
where
\begin{eqnarray} \label{eq34}
\begin{split}
  |\tilde{\phi}_{0}^{3,3}\rangle = & \frac{1}{3}(|000000\rangle +|111111\rangle +|222222\rangle\\&
                                                +|012012\rangle +|120120\rangle +|201201\rangle\\&
                                                +|021021\rangle +|102102\rangle +|210210\rangle)_{A_1B_1C_1A_2B_2C_2},
\end{split}
\end{eqnarray}
\begin{eqnarray} \label{eq35}
\begin{split}
 |\tilde{\phi}_{1}^{3,3}\rangle = & \frac{1}{3}(|001001\rangle +|112112\rangle +|220220\rangle\\&
                                               +|010010\rangle +|121121\rangle +|202202\rangle\\&
                                               +|022022\rangle +|100100\rangle +|211211\rangle)_{A_1B_1C_1A_2B_2C_2},
\end{split}
\end{eqnarray}
\begin{eqnarray} \label{eq36}
\begin{split}
  |\tilde{\phi}_{2}^{3,3}\rangle = & \frac{1}{3 }(|002002\rangle +|110110\rangle +|221221\rangle\\&
                                                 +|011011\rangle +|122122\rangle +|200200\rangle\\&
                                                 +|020020\rangle +|101101\rangle +|212212\rangle)_{A_1B_1C_1A_2B_2C_2},
\end{split}
\end{eqnarray}

Next, as shown in Figure~ \ref{Fig4}, after the six photons suffer from $\mathcal{F}_3^{-1}$s and $\mathcal{F}_3$, project measurements, and classical feed-forward operations in succession (see \textbf{Table \ref{Tab2}}), we can obtain the desired state $|\psi_{0,0}^{3,3}\rangle_{A_1B_1C_1}$ with an improved fidelity of 
$\frac{\tilde{F}_0^2}{\tilde{F}_0^2 + \tilde{F}_1^2 + \tilde{F}_2^2} > \tilde{F}_0$ if and only if $\tilde{F}_0 > \frac{1}{3}$, and the detailed proof process is presented in Appendix \ref{secD}.


\subsection{EPP for $n$-qudit GHZ state with phase-flip error channels}\label{sec3-2}

Our program can be extended to distill the arbitrary $d$-dimensional $n$-partite GHZ state $|\psi_{0,0}^{d,n}\rangle_{A_1B_1\cdots N_1}$ with qudit-based phase-flip error $Z_{d_i}$. 
Here $Z_{d_i}$ can be expressed as
\begin{eqnarray} \label{eq37}
  \begin{split}
   Z_{d_i} = \sum_{j=0}^{d-1} e^{\text{i}\frac{2\pi}{d}((j+i) \bmod d)}| j \rangle  \langle j |, 
  \end{split}
\end{eqnarray}
where $i=0,1,\cdots, (d-1)$. 
These errors make $|\psi_{0,0}^{d,n}\rangle_{A_1B_1\cdots N_1}$ become
\begin{eqnarray} \label{eq38}
 \bar{\rho}_{A_1B_1\cdots N_1}=\sum_{l=0}^{d-1} \bar{F}_{l}|\psi_{l,0}^{d,n}\rangle_{A_1B_1\cdots N_1}\langle \psi_{l,0}^{d,n}|.
\end{eqnarray}
Based on Eq. \eqref{eq14}, state $|\psi_{l,0}^{d,n}\rangle_{A_1B_1\cdots N_1}$ can be reedited as
\begin{eqnarray} \label{eq39}
|\psi_{l,0}^{d,n}\rangle_{A_1B_1\cdots N_1}=\frac{1}{\sqrt d}\sum_{j=0}^{d-1}e^{\mathrm{i}\frac{2\pi}{d}lj} (|j\rangle^{\otimes n})_{A_1 B_1 \cdots N_1}.
\end{eqnarray}

For correcting the qudit-based phase-flip error in $n$-qudit GHZ state, we should follow the same arrangement as that made for three-qutrit GHZ state (see Sec. \ref{sec3-1}). 
Here the qutrit-based QND is replaced with the qudit-based QNDs shown in Figure~ \ref{Fig3}, and the number of the qudit-based QND equipments is increased from 3 to $n$. 
The desired output state $|\psi_{0,0}^{d,n}\rangle_{A_1B_1\cdots N_1}$ can be obtained with an enhanced fidelity of 
$\frac{\bar{F}_0^2}{\sum_{l=0}^{d-1} \bar{F}_l^2} > \bar{F}_0$, with $\bar{F}_0 > \frac{1}{d}$. 
The detailed proof process can be found in Appendix  \ref{secE}.

\section{Discussion and Summary}\label{sec4}

High-dimensional multipartite quantum entanglement represents a significant advancement over the two-dimensional bipartite counterparts. 
In 2007, Cheong et al. \cite{high-dimensional-EPPs2} proposed an interesting EPP for three-qutrit generalized GHZ state via generalized CNOT gates, and the trit-flip and phase errors are corrected. 
Generalized CNOT gate has been experimentally realized in superconducting circuits \cite{superconduct} and hybrid photonic systems \cite{hybrid}, and generalized multi-photon CNOT gate is  rarely investigated.
In 2009, Sheng et al. \cite{Kerr13} presented EPPs for three-qubit GHZ state with qubit-flip and phase-flip errors, and QNDs are employed to bypass the experimental challenge of CNOT gate.

We proposed high-dimensional EPPs for extracting a subset of high quality arbitrary $d$-dimensional $n$-partite GHZ state from a set of less-entangled $n$-qudit GHZ states with qudit-flip or qudit-based phase-flip errors. 
In our schemes, qudits can be encoded in single-photon systems using $d$ orthogonal optical modes in a variety of DOF, such as spatial modes, OAMs, optical frequencies, and time-bins. 
It is known that the spatial single-qudit operations can be well achieved by employing the Mach-Zehnder Interferometers and phase shifters \cite{F1,F2}.


\begin{table}[htbp]
  \centering 
  \caption{Comparison between the high-dimensional multipartite EPP in Ref.~\cite{high-dimensional-EPPs2} and the present work. The fidelity threshold of our scheme for the general $(d, n)$ scenario is $\frac{1}{d^{n-1}}$ (derived in Appendix~\ref{secC}). Specifically, for the $(3, 3)$ scenario, the fidelity threshold of our scheme is $\frac{1}{9} \approx 0.111$, and the counterpart in Ref.~\cite{high-dimensional-EPPs2} is $0.2305$. Furthermore, our scheme is extended to arbitrary $(d, n)$ scenario with $d \geq 3$ and $n \geq 3$.}
 \label{Tab4}
 \begin{tabular}{ccc}
 \hline  \hline 

  Name                                           & Ref.~\cite{high-dimensional-EPPs2}   & Our work \\
   \hline  
  State                                          &  GHZ                                 &  GHZ \\
  Noise                                          &  qudit-flip, phase-flip              & qudit-flip, phase-flip\\
  Trick                                          & generalized CNOT                     & cross-Kerr nonlinearity\\
  Source for 3-qutrit GHZ case                   & 3 generalized CNOTs                  & 2 QND \\
  $d$-level                                      & $d=3$                                & $d\geq 3$ \\
  $n$-partite                                    & $n\geq 3$                            & $n\geq 3$\\
  Fidelity Threshold (3-level 3-partite)         & 0.2305                               & $\frac{1}{9}$\\
  Fidelity Threshold ($d$-level $n$-partite)     & N/A                                  & $\frac{1}{d^{n-1}}$ \\
  Iterate 2 rounds                               & 48 copies                            & 8 input copies\\

\hline \hline 
\end{tabular}
\end{table}

In our proposed EPP, the fidelity of the output can be further improved by iterating the EPP processes. 
For the sake of clarity, we take the EPP for correcting one qutrit-flip as an example. 
As shown in Sec. \ref{sec2-2}, in our purification program,  Alice, Bob, and Charlie only pick up the phase shift $(0_A,0_B,0_C)$ and discard the other scenarios. 
In this way, the photons kept are in the state
\begin{eqnarray} \label{eq40}
  \widehat{\rho}_{A_1B_1C_1A_2B_2C_2}= \frac{1}{3} (F_0^2 |\psi _{0,0}\rangle_{A_1B_1C_1A_2B_2C_2}\langle\psi_{0,0}| 
                                                 +  F_1^2 |\psi _{1,0}\rangle_{A_1B_1C_1A_2B_2C_2}\langle\psi_{1,0}|  
                                                 +  F_2^2 |\psi _{2,0}\rangle_{A_1B_1C_1A_2B_2C_2}\langle\psi_{2,0}|).
\end{eqnarray}  
Here $|\psi _{0,0}\rangle$, $|\psi _{1,0}\rangle$, and $|\psi _{2,0}\rangle$ are given in Eq. \eqref{eq3}.
In fact, the protocol can not correct the qutrit-flip error completely, but it can be further purified by iterating the EPP process discussed in Sec. \ref{sec2-2}.   
Hence another additional state $\widehat{\rho'}$ should be employed. Here
\begin{eqnarray} \label{eq41}
  \widehat{\rho'}_{A_3B_3C_3A_4B_4C_4}= \frac{1}{3} (F_0^2 |\psi _{0,0}\rangle_{A_3B_3C_3A_4B_4C_4}\langle\psi_{0,0}| 
                                                   + F_1^2 |\psi _{1,0}\rangle_{A_3B_3C_3A_4B_4C_4}\langle\psi_{1,0}|  
                                                   + F_2^2 |\psi _{2,0}\rangle_{A_3B_3C_3A_4B_4C_4}\langle\psi_{2,0}|).
\end{eqnarray} 
In this way, after iterating the EPP process $t$ rounds,  the desired state $|\psi_{0,0}^{3,3}\rangle$ can be obtained with a further improved fidelity  
\begin{eqnarray} \label{eq42}
    F'_t=\frac{F_0^{2^t}}{F_0^{2^t} + F_1^{2^t} + F_2^{2^t}}.
\end{eqnarray} 

\begin{figure}[htbp]
 \centering
 \begin{subfigure}[b]{0.45\textwidth}
        \centering
        \includegraphics[width=\textwidth]{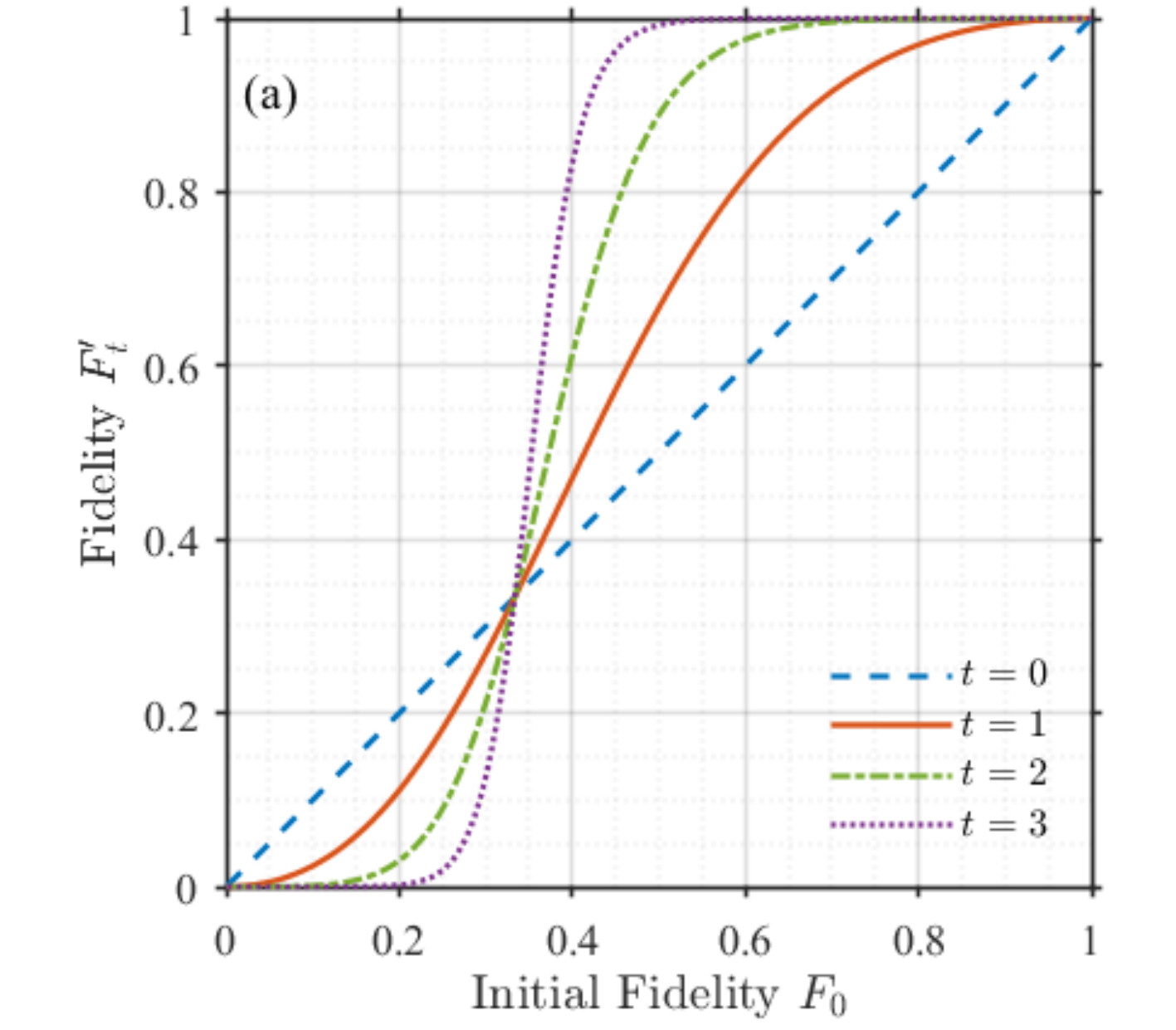}
        \phantomcaption
        \label{Fig5(a)}
 \end{subfigure}
    \hfill 
    \begin{subfigure}[b]{0.45\textwidth}
        \centering
        \includegraphics[width=\textwidth]{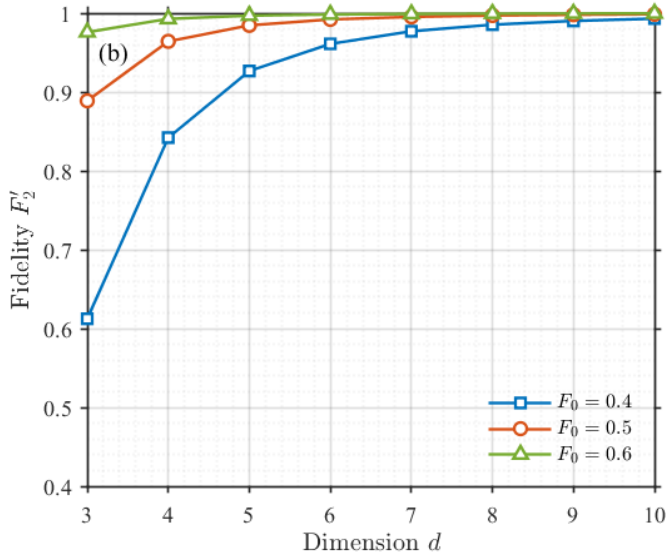}
        \phantomcaption
        \label{Fig5(b)}
    \end{subfigure}
 \caption{Fidelity of the EPP protocol for correcting  qutrit-flip errors. 
    (a) The fidelity $F'_t$ as a function of input fidelity $F_0$, where $d=3$ and the iterative rounds $t=0, 1, 2, 3$.       
    (b) The fidelity $F'_2$ as a function of the dimension $d$, where $t=2$ and $F_0 = 0.4, 0.5, 0.6$.
    }
 \label{Fig5}
\end{figure}

\textbf{Figure \ref{Fig5}} shows the output fidelities versus the input fidelity, versus the iteration rounds, and versus the dimension. 
Our results show that the fidelity thresholds of the proposed EPP for correcting qutrit-flip errors is $F_0=\frac{1}{3}$, which agree with the result shown in Appendix \ref{secA} very well.
The fidelity increase with the input fidelity $F_0$, the number of iteration rounds $t$, and the dimension $d$. 
Fidelity of the output can be asymptotically improved to unity by iterating the EPP process.

Kerr nonlinearity is a key ingredients of our schemes. \textbf{Table \ref{Tab4}} shows the comparison between the CNOT-based EPP \cite{high-dimensional-EPPs2} and the proposed EPP. 
It is known that the nonlinearity of the natural Kerr media is extremely small, and  giant Kerr nonlinearity is an experimental challenge.
There have been numerous theoretical and experimental studies to amplify the strength of cross-Kerr nonlinearity. 
Utilizing electromagnetically induced transparency \cite{EIT1}, the natural cross-Kerr nonlinearity can be enhanced from  $\theta\sim 10^{-18}$ to $\theta\sim10^{-2}$.
In 2016, Tiarks et al. \cite{Tiarks2016SciAdv} experimentally demonstrated that phase shift $\pi$ created by a single-photon pulse in Rydberg media. Moreover, quantum dot spin mediate has also been employed to build QNDs.

In conclusion, we have presented scalable EPP protocols for arbitrary $d$-dimensional $n$-partite GHZ states, addressing both qudit-flip and phase-flip errors. 
Beyond conventional qubit-based EPPs, our EPPs work in high-dimensional quantum systems, providing a viable pathway for high-dimensional quantum processing.
It is known that CNOT gates are necessary in Murao's EPP for qubit multipartite entangled state.
In our QND EPPs, the roles of the generalized CNOT gate and photon-number-resolving detectors are fulfilled by QNDs with cross-Kerr nonlinearities. 
By iterating the EPP process, fidelity of the output can be asymptotically improved to unity. 
Moreover, the fidelity thresholds for the proposed EPPs are developed.
These features suggest that our scheme may offer an alternative way for distributing high-fidelity high-dimensional entanglement in future quantum networks.

\medskip

\textbf{Acknowledgements} \par 
This work is supported by the National Natural Science Foundation of China under Grant No. 62371038; 
the Beijing Natural Science Foundation under Grant No. 4252006.


\medskip
\appendix

\section{The fidelity threshold of the EPP for three-qutrit GHZ state with one qutrit-flip errors channel shown in Sec. \ref{sec2-2}} \label{secA}

As shown in Sec. \ref{sec2-2}, the success of EPP for three-qutrit GHZ state with one qutrit-flip errors channel indicates that
\begin{eqnarray} \label{eqA1}
\begin{split}
    \frac{F_0^2}{F_0^2 + F_1^2 + F_2^2} > F_0. 
\end{split}
\end{eqnarray}
Here $F_0, F_1, F_2 \in (0,1)$ and $F_0 + F_1 + F_2=1$. 
By dividing both sides of Eq. \eqref{eqA1} by $F_0$, we obtain 
\begin{eqnarray} \label{eqA2}
\begin{split}
    \frac{F_0}{F_0^2 + F_1^2 + F_2^2} > 1.
\end{split}
\end{eqnarray}
Eq. \eqref{eqA2} can be written as
\begin{eqnarray} \label{eqA3}
\begin{split}
    F_1^2 + F_2^2 < F_0 - F_0^2.
\end{split}
\end{eqnarray}
Based on the Cauchy-Schwarz inequality, we have
\begin{eqnarray} \label{eqA4}
\begin{split}
    F_1^2 + F_2^2 \ge  \frac{(1 - F_0)^2}{2}.
\end{split}
\end{eqnarray}
Substituting Eq. \eqref{eqA4} back into Eq. \eqref{eqA3}, and rewriting the expressions we obtain
\begin{eqnarray} \label{eqA5}
\begin{split}
    \frac{(1 - F_0)^2}{2} < F_0 - F_0^2.
\end{split}
\end{eqnarray}
Eq. \eqref{eqA5} can be simplified as
\begin{eqnarray} \label{eqA6}
\begin{split}
    3F_0^2 - 4F_0 + 1 < 0.
\end{split}
\end{eqnarray}
Hence, one can calculate that the valid range for the parameter $F_0$ is
\begin{eqnarray} \label{eqA7}
\begin{split}
    F_0 \in ( \frac{1}{3}, 1 ). 
\end{split}
\end{eqnarray}


\section{The fidelity threshold of the EPP for $n$-qudit GHZ state with one qudit-flip errors channel shown in Sec. \ref{sec2-3}}  \label{secB}

As shown in Sec. \ref{sec2-3}, the success of the EPP for $n$-qudit GHZ state with one qudit-flip errors channel  means that
\begin{eqnarray} \label{eqB1}
\begin{split}
    \frac{\dot{F}_{0}^2}{\sum_{u=0}^{d-1} \dot{F}_u^2} > \dot{F}_0.
\end{split}
\end{eqnarray}
Eq. \eqref{eqB1} can be rewritten as
\begin{eqnarray} \label{eqB2}
\begin{split}
    \sum_{{u}=1}^{d-1} \dot{F}_u^2  < \dot{F}_0 - \dot{F}_0^2.
\end{split}
\end{eqnarray}
Based on the Cauchy-Schwarz inequality, we have
\begin{eqnarray} \label{eqB3}
\begin{split}
    \sum_{u=1}^{d-1} \dot{F}_u^2 \ge  \frac{(1 - \dot{F}_0)^2}{d-1}.
\end{split}
\end{eqnarray}
Substituting Eq. \eqref{eqB3} back into Eq. \eqref{eqB2},  we obtain
\begin{eqnarray} \label{eqB4}
\begin{split}
    (d\dot{F}_0 - 1)(\dot{F}_0 - 1) < 0.
\end{split}
\end{eqnarray}
The valid range for the parameter $\dot{F}_0$ can be calculated as
\begin{eqnarray} \label{eqB5}
\begin{split}
     \dot{F}_0 \in ( \frac{1}{d}, 1 ).
\end{split}
\end{eqnarray}

\section{The fidelity threshold of the EPP for $n$-qudit GHZ state with $n$ nontrivial qudit-flip errors channels shown in Sec. \ref{sec2-4}} \label{secC}

As shown in Sec. \ref{sec2-4}, the success of EPP for $n$-qudit GHZ state with $n$ nontrivial qudit-flip errors channels indicates that
\begin{eqnarray} \label{eqC1}
\begin{split}
    \frac{\ddot{F}_0^2}{\sum_{u=0}^{d^{n-1}-1} \ddot{F}_u^2} > \ddot{F}_0.
\end{split}
\end{eqnarray}
This can be recast as
\begin{eqnarray} \label{eqC2}
\begin{split}
    \sum_{u=1}^{d^{n-1}-1} \ddot{F}_u^2 < \ddot{F}_0 - \ddot{F}_0^2.
\end{split}
\end{eqnarray}
Using the Cauchy-Schwarz inequality,
\begin{eqnarray} \label{eqC3}
\begin{split}
    \sum_{u=1}^{d^{n-1}-1} \ddot{F}_u^2 \ge \frac{(1 - \ddot{F}_0)^2}{d^{n-1}- 1}.
\end{split}
\end{eqnarray}
Substituting Eq.~\eqref{eqC3} back into Eq.~\eqref{eqC2} yields
\begin{eqnarray} \label{eqC4}
\begin{split}
    \bigl(d^{n-1} \ddot{F}_0 - 1\bigr)(\ddot{F}_0 - 1) < 0.
\end{split}
\end{eqnarray}
Therefore, the valid range of $\ddot{F}_0$ is
\begin{eqnarray} \label{eqC5}
\begin{split}
    \ddot{F}_0 \in ( \frac{1}{d^{n-1}}, 1 ).
\end{split}
\end{eqnarray}

\section{The fidelity threshold of the EPP for three-qutrit GHZ states with phase-flip errors channels shown in Sec. \ref{sec3-1}}  \label{secD}

As shown in Sec. \ref{sec3-1}, the success of EPP for 3-qutrit GHZ state with phase-flip errors channels indicates that
\begin{eqnarray} \label{eqD1}
\begin{split}
    \frac{\tilde{F}_0^2}{\tilde{F}_0^2 + \tilde{F}_1^2 + \tilde{F}_2^2} > \tilde{F}_0.
\end{split}
\end{eqnarray}

Following the same principle shown in Appendix \ref{secA}, we can find that the valid range of $\tilde{F}_0$ is
\begin{eqnarray} \label{eqD2}
\begin{split}
    \tilde{F}_0 \in ( \frac{1}{3}, 1 ). 
\end{split}
\end{eqnarray}

\section{The fidelity threshold of the EPP for $n$-qudit GHZ state with phase-flip errors channel shown in Sec. \ref{sec3-2}} \label{secE}

As shown in Sec. \ref{sec3-2}, the fidelity  of the EPP addressing all phase-flip errors means that
\begin{eqnarray} \label{eqE1}
\begin{split}
    \frac{\bar{F}_0^2}{\sum_{l=0}^{d-1} \bar{F}_l^2} > \bar{F}_0. 
\end{split}
\end{eqnarray}

Following the same principle shown in Appendix \ref{secB}, we can find that the valid range of $\bar{F}_0$ is
\begin{eqnarray} \label{eqE2}
\begin{split}
    \bar{F}_0 \in ( \frac{1}{d}, 1 ). 
\end{split}
\end{eqnarray}




\end{document}